\documentclass[%
12pt,T
 oneocolumn,
 nofootinbib,
 amsmath,amssymb,
 aps,
floatfix,
unsortedaddress
]{revtex4-2}

\usepackage{graphicx,color}
\usepackage{subfig}
\usepackage{dcolumn}% Align table columns on decimal point
\usepackage{bm}% bold math
\usepackage{graphicx,color}
\usepackage{float}
\usepackage{romannum}

\usepackage{amsmath}
\usepackage{mathtools}

\usepackage{cancel}
\usepackage[pdfpagelabels, pdfencoding=auto, psdextra]{hyperref}
\hypersetup{%
 pdfsubject=Paper,
 pdfkeywords={nuclear physics} {Pionless EFT} {Two-Nucleon} {Lattice} {L\"usher formula},
 unicode = true,
 breaklinks = true,
 colorlinks = true,
 linkcolor = blue,
 citecolor = blue,
 menucolor = blue,
 citecolor = blue,
 urlcolor = blue
}

\graphicspath{{./}}
\begin{document}
	\pagenumbering{arabic}

\title{  $ _{\Xi}^{5}H $ hypernuclei by folding the state-of-the-art $ \Xi N $ interactions}
\author{Faisal Etminan}
 \email{fetminan@birjand.ac.ir}
\affiliation{
 Department of Physics, Faculty of Sciences, University of Birjand, Birjand 97175-615, Iran
}%
\affiliation{ Interdisciplinary Theoretical and Mathematical Sciences Program (iTHEMS), RIKEN, Wako 351-0198, Japan}

\date{\today}% 

\begin{abstract}
	
			I examined  a phenomenological Nijmegen and a first principles HAL QCD $\Xi N$ potentials
			 to study $\alpha\Xi$ ineractions. 
			 A Woods-Saxon type form for $\alpha\Xi$ potential in the single-folding potential approach is derived
			 by using the spin- and isospin averaged $\Xi N$ interactions.
			 The possibility of resonance or bound state is searched and accordingly,  
			 the low energy scattering phase shift parameters of $\alpha\Xi$ are calculated.
			 The numerical results  show that 
			 even though two $ \Xi N $ potentials have significantly dissimilar isospin (I) and spin (S) components,
			 $ _{\Xi}^{5}H $   could be only a Coulomb-assisted resonance state that
			 appears about  $ 0.5 $ MeV below the threshold of $\alpha+\Xi^{-} $ for both model of potentials.
\end{abstract}

%\begin{keyword}
% HAL QCD $ \Xi N $ potential, Nijmegen $ \Xi N $ potential,
% single-folding potential approach, $\alpha\Xi$ Woods-Saxon type potential, 
% Coulomb-assisted resonance state
%\end{keyword}

\maketitle
%%%%%%%%%%%%%%%%%%
\section{Introduction} \label{sec:intro}

Although there is a lot of interest in knowing about hyperon interactions, the underlying $ \Xi N $ interaction  is facing many  uncertainties. The main reason is the lack of experimental data. In this regard, many researches have been conducted recently, both experimentally and theoretically, such as
 high resolution gamma-ray experiments~\cite{Tamura2000}, KISO event~\cite{Nakazawa2015}, femtoscopic measurements of pp, pA and AA collisions in the ALICE and STAR
 experiments~\cite{PhysRevLett.123.112002}, femtoscopic study of $ \Xi N $ correlation function
  in heavy-ion collisions ~\cite{HATSUDA2017856,PhysRevC.105.014915}  and
 advanced few-body theoretical 
methods~\cite{10.1143/ptp.117.251,hiyama2000,hiyama2008prc,hiyama2014,hiyama2018annurev,Hiyama2020,Garcilazo2016prc,Garcilazo2020cpc,Garcilazo2020cpc2,etminan2024prc},
 Jacobi no-core shell model (J-NCSM)~\cite{le2021} and 
 lattice QCD
 calculations~\cite{Sasaki2014ptep,sasaki2020}.

Hiyama et al. in Ref.~\cite{hiyama2008prc} studied  $ _{\Xi}^{5}H $ hypernuclei by using two types of $  \Xi N$ interactions, the Nijmegen Hard-Core model D (ND)~\cite{PhysRevD.15.2547} and the Extended
Soft-Core model (ESC04) ~\cite{rijken2006extendedsoftcore},
it was found to be a Coulomb-assisted bound state, in other words, no nuclear-bound state appeared.
 While  the existence of bound  $ \Xi $
states in systems with $ A = 4-7 $ baryons based on the  chiral $  \Xi N$ interactions  are investigated in Ref.~\cite{le2021} and a clearly bound state about $ 2.16 \pm 0.1 $ MeV is reported for 
$ \alpha \Xi\left[_{\Xi}^{5}H\left(J^{\pi}=\frac{1}{2}^{+},T=\frac{1}{2}\right)\right] $. 
 Also, Friedman and Gal~\cite{FRIEDMAN2021136555} by utilizing an optical potential concluded same result for binding energy of
 $ _{\Xi}^{5}H $ that was $ 2.0 $ MeV.  
 Furthermore, Myint and Akaishi~\cite{10.1143/ptp.117.251}  by using Nijmegen model-D potential~\cite{PhysRevD.15.2547} 
 reported $ 1.7 $ MeV  for binding energy of this system, where  authors   emphasized that the large part binding energy is due to $\alpha+\ensuremath{\Xi^{-}}$ Coulomb interaction.

   The $ NN \Xi $  and $ NNN \Xi $ system studied in Ref.~\cite{Hiyama2020} by employing  two modern $  \Xi N $ interactions:
       a phenomenological Nijmegen $  \Xi N $ potential (ESC08c) that is established on the meson  exchanges~\cite{nagels2015extendedsoftcore} and HAL QCD  $ \Xi N $ potential (HAL QCD) which is based on  first principle lattice QCD simulations~\cite{sasaki2020}. In Ref.~\cite{Hiyama2020} systems with more particles have not been investigated.
  Therefore, by considering the above results it is necessary to study $ _{\Xi}^{5}H $ by using both state-of-the-art 
  ESC08c and HAL QCD 
    $  \Xi N $ potential where the latter is the most consistent potentials with the LHC ALICE data~\cite{Fabbietti:2020bfg,alice2020unveiling}. Recently, the HAL QCD Collaboration published the first lattice QCD
 simulations of the $\Xi N$ potential ~\cite{sasaki2020}.
 The simulation was performed by $\left(2+1\right)$-flavor with quark
 masses near the physical point $m_{\pi}\simeq146.4$ MeV and $m_{K}\simeq525$
 MeV on a large lattice space- time volume $\simeq\left(8.1\:{\textrm{fm}}\right)^{4}$.
  In addition, the dibaryons
 systems with multiple strangeness such as $\Lambda\Lambda,\Xi N$
 ~\cite{sasaki2020}, $\Omega N$ ~\cite{etminan2014,Iritani2019prb}
 and $\Omega\Omega$~\cite{Gongyo2018}
 have been calculated by HAL QCD method~\cite{Ishii2007,ishii2012,aoki2013}.

 In this work, a Woods-Saxon type form for $ \alpha+\Xi $ system is obtained by using single-folding potential (SFP) method. Since the $ \alpha $-cluster has low compressibility property, in this model it is supposed that both particles $\Xi $ and $\alpha$  move in an effective $\alpha \Xi$ potential.   
   The effective $ \alpha+\Xi $ nuclear potential is approximated by the single-folding of nucleon density in
 the $\alpha$-particle and spin- and isospin averaged $ \Xi N $ potential between $ \Xi $  and a nucleon~\cite{Satchler1979,10.1143/ptp.117.251,Miyamoto2018,Etminan:2019gds}. 
 
 After that, the resultant $\alpha \Xi$ potential is fitted to a continues analytical Woods-Saxon type function. Then the Schr\"{o}dinger equation is solved in the infinite volume by using this fitted function as input to obtain binding energy and  relevant scattering observables
 from the asymptotic behavior of the wave function. 
 
 It should be noted here that 
  the SFP models reduced a  five-body problem to an effective two-body problem, 
 this can be computational advantage but this approximation may miss some part of the physics of the problem,
 e.g. the  obtained potential might be underestimated. So to treat this issue as much as possible,
 here the SFP method is benchmarked and tuned by employing  ESC08c  $\Xi N$ potential only in  
  $ \ensuremath{^{3}S_{1}}\left(I=0\right) $ ~\cite{nagels2015extendedsoftcore} channel as input to reproduce the parameters of common phenomenological Dover-Gal (DG)  $\alpha \Xi$ potential~\cite{dover1983,Etminan:2019gds}.  And, this model is accurate just for the low energy 
  features of the $\alpha+\Xi$ system, because it is extracted from the low
  energy $\Xi N$ interactions.   
There are  enormous applications of $\alpha \Xi$ potential in few-body methods based on $\alpha$-cluster models such as 3-, 4- and 5-body 
Gaussian  expansion~\cite{hiyama2018annurev}, J-NCSM~\cite{le2021} and Faddeev~\cite{Filikhin2008,Garcilazo2016prc,Miyagawa2021krh,Etminan:2022bon}.

The paper is organized as follows.
In Sec.~\ref{sec:folding-Model},  I provide approximately reliable expressions for the effective $\Xi N$ interactions that is defined by averaging on the spin- and isospin components for both ESC08c and HAL QCD $\Xi N$ potentials. Also, the single-folding potential approach is described briefly. In Sec.~\ref{sec:result},  the numerical results and discussions are presented.
 The summary and final conclusions  are given in Sec.~\ref{sec:Summary-and-conclusions}. 
%%%%%%%%%
\section{$ \Xi N $ interactions  and single-folding potential approach } \label{sec:folding-Model}
Here, the ESC08c and the HAL QCD $\Xi N$ potentials that are used to found  effective potentials of $ \alpha +\Xi $ systems are described. And a brief description of the SFP model~\cite{Satchler1979,Etminan:2019gds} is given.

The  ESC08c Nijmegen $\Xi N$ potential function consists
of local central Yukawa-type potentials with attractive and repulsive terms~\cite{nagels2015extendedsoftcore,Garcilazo2016prc}, 
\begin{equation}
	V_{\Xi N}^{ESC08c}\left(r\right)=-A\frac{\exp\left(-\mu_{A}r\right)}{r}+B\frac{\exp\left(-\mu_{B}r\right)}{r}.
	\label{eq:pot-XiN}
\end{equation}
where the low-energy data and the parameters of these models are given in Table I of Ref.~\cite{Garcilazo2016prc}. 

For  $ \Xi N $ HAL QCD potential, the concrete parametrizations, are taken
straight from Ref.~\cite{sasaki2020}
at the imaginary-time slices $ t/a=11,12,13 $ where $ a=0.0846 $ fm is the lattice spacing. Sasaki et al. in Ref.~\cite{sasaki2020} 
 clearly treated $ \Lambda\Lambda-\Xi N $ interactions  by the HAL QCD coupled-channel formalism~\cite{aoki2013,etminan2024prd}.
The S-wave $ \Xi N $ interactions is classified in four channels $^{11}S_{0},^{31}S_{0},^{13}S_{1}$
and $^{33}S_{1}$. Here same as Ref.~\cite{sasaki2020} the spectroscopic notation $^{2I+1,2S+1}S_{J}$ is employed
where $ I, S $ and $ J $ indicate the total isospin, the total spin,
and the total angular momentum, respectively. 
%Coupled-channel contribution from the lower channel ($ \Lambda\Lambda $ in $ ^{11}S_{0} $) is clearly treated by the HAL QCD coupled-channel formalism~\cite{aoki2013,etminan2024prd}. While from higher channels ($ \Xi N-\Lambda\Sigma-\Sigma\Sigma $ in $ ^{33}S_{1} $)  are renormalized into a $ \Xi N-\Xi N $ effective central potential by including a Gaussian function as, $ -233\exp\left(-r^{2}\right) $~\cite{Hiyama2020}.

For phenomenological applications, the lattice QCD potential was fitted
by analytic functional forms composed of multiple Gaussian and Yukawa
functions. The Gauss functions set out the short range part of the
potential and  Yukawa functions describe the meson exchange picture
at medium to long range distances
\begin{eqnarray}
	V_{\Xi N}\left(r\right) & =\sum_{i=1}^{3} & \alpha_{i}e^{-r^{2}/\beta_{i}^{2}}+\lambda_{2}\left[\mathcal{Y}\left(\rho_{2},m_{\pi},r\right)\right]^{2}+\lambda_{1}\mathcal{Y}\left(\rho_{1},m_{\pi},r\right),
\end{eqnarray}
where the values of the parameters $ \alpha_{1,2,3},\beta_{1,2,3},\lambda_{1,2} $ and
$ \rho_{1,2} $ are given in Table 4 of Ref~\cite{sasaki2020}.
$ m_{\pi}\simeq146 $ MeV is the pion mass that was measured on the
lattice. I have also presented results by considering
the physical value of the pion mass $ m_{\pi}\simeq138 $ MeV. The form
factor $\mathcal{Y}$ defines Yukawa function 
\begin{equation}
	\mathcal{Y}\left(\rho,m,r\right)\equiv\left(1-e^{-\frac{r^{2}}{\rho^{2}}}\right)\frac{e^{-mr}}{r}.
\end{equation}
Since the $\alpha\Xi^{-}\left(_{\Xi^{-}}^{5}H\right)$ system required
to show a clear core-$\Xi$ structure, I employ approximately reliable
expressions for spin and isospin dependence of $\Xi N$ interaction
to $s$-shell (the core nucleons and the $\Xi$ are in S-wave states
).
Therefore in these approximation, the effective $\Xi N$ interactions
can be obtained by averaging on the spin- and isospin components~\cite{hiyama2008prc,le2021} 
\begin{equation}
	\bar{V}_{\Xi N}\simeq\frac{\left(V_{\Xi N}^{^{11}S_{0}}+3V_{\Xi N}^{^{31}S_{0}}+3V_{\Xi N}^{^{13}S_{1}}+9V_{\Xi N}^{^{33}S_{1}}\right)}{16}.
\end{equation}

\begin{figure*}[hbt!]
	\centering
	\includegraphics[scale=0.64]{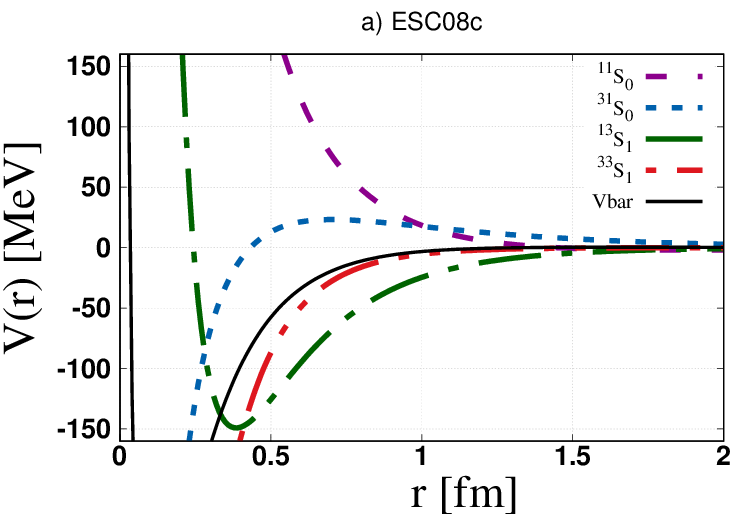} \includegraphics[scale=0.64]{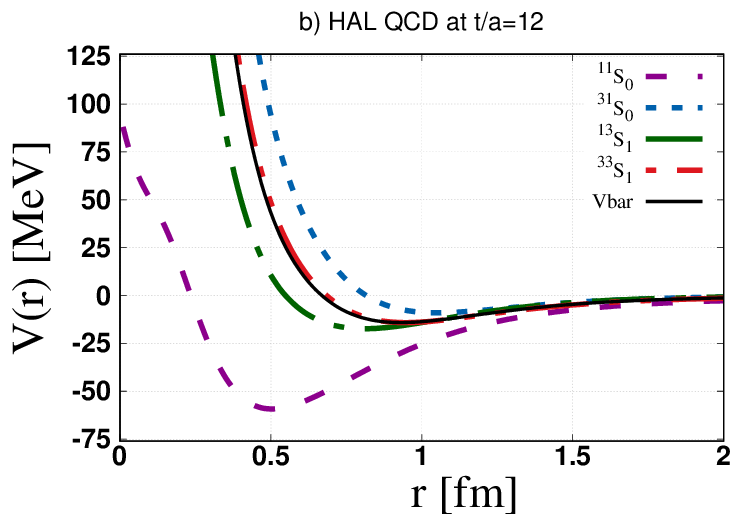}
	\caption{	(a) ESC08c  and (b) HAL QCD $ \left(t/a=12\right) $ $\Xi N $ potentials in $^{11}S_{0},^{31}S_{0},^{13}S_{1}$
		and $^{33}S_{1}$ channels. The corresponding scattering phase shifts for each channel are given in Ref.~\cite{Hiyama2020}. The solid black line, which is labeled by Vbar, shows the  spin- and isospin 	 averaged potential, $\bar{V}_{\Xi N} $ in Eq.~\eqref{eq:pot-XiN}.	\label{fig:NXi_pot}}
\end{figure*}

In Fig.~\ref{fig:NXi_pot}, I compare $\Xi N$ potential for (a) the ESC08c model and (b) the HAL QCD at
$ t/a = 12 $~\cite{sasaki2020}. The statistical errors are not shown in Fig.~\ref{fig:NXi_pot}(b), but are considered in my calculations. 
%I should mention that in my model of calculation only ESC08c Nijmegen $\Xi N$ potential in the $ ^{13}S_{1} $ channel reproduced the parameters of  phenomenological Dover-Gal (DG)  $\alpha \Xi$ potential~\cite{dover1983,Etminan:2019gds}.
%%%%%%%%%%%%%%%%
%%%%%%%%%%%%%%%%
Fig.~\ref{fig:NXi_pot} reveals qualitative difference between (a) and (b).
Therefore it is interesting to find out how these differences are embodied in the low energy properties of $ _{\Xi}^{5}H $.  

The effective $ \alpha+\Xi $ nuclear potential is approximated by the single-folding of nucleon density $\rho\left(r^{\prime}\right)$ in
the $\alpha$-particle and spin- and isospin averaged $\Xi N$ potential, $\bar{V}_{\Xi N}\left(\left|\textbf{r}-\textbf{r}^{\prime}\right|\right)$ between the $\Xi$ particle at $\textbf{r}^{\prime}$
and the nucleon at $\textbf{r}$~\cite{Satchler1979,Miyamoto2018,Etminan:2019gds}.
The $\alpha \Xi$ potential is defined as
\begin{equation}
	V_{\alpha\Xi}\left(r\right)=\int\rho\left(r^{\prime}\right)\bar{V}_{\Xi N}\left(\left|\textbf{r}-\textbf{r}^{\prime}\right|\right)d\textbf{r}^{\prime},\label{eq:V_alfaOmega}
\end{equation}		
where $\rho\left(r^{\prime}\right)$ defined the nucleon density function in
 $\alpha$-particle at a distance $\textbf{r}^{\prime}$ from its
center-of-mass, and it can be chosen as follows~\cite{Akaishi1986},
\begin{equation}
	\rho\left(r^{\prime}\right)=4\left(\frac{4\beta}{3\pi}\right)^{3/2}\exp\left(-\frac{4}{3}\beta r^{\prime2}\right).\label{eq:nucleon-density}
\end{equation}
The integration in Eq.~\eqref{eq:V_alfaOmega} is over all space as permitted
by $\rho\left(r^{\prime}\right)$. Normalization conditions require that
\begin{equation}
	\int\rho\left(r^{\prime}\right)d\textbf{r}^{\prime}=4\left(\frac{4\beta}{3\pi}\right)^{3/2}\intop_{0}^{\infty}\mathop{\exp\left(-\frac{4}{3}\beta r^{\prime2}\right)4\pi r^{\prime2}dr^{\prime}}=4,
\end{equation}
here,  $\beta$ is a constant and it could be determined by the rms radius of $\textrm{\ensuremath{^{4}}He}$~\cite{Akaishi1986}, $ 	\textrm{\ensuremath{r_{r.m.s}}}=\frac{3}{\sqrt{8\beta}}=1.47\:\textrm{fm}. $ 

Finally,  a hard-sphere model of Coulomb interaction is employed in my calculations   as~\cite{Etminan:2022bon}
\begin{equation}
	V_{Coul}\left(r\right)=-2\alpha_{f} \times\begin{cases}
		\frac{1}{r_{Coul}}\left(\frac{3}{2}-\frac{r^{2}}{2r_{Coul}^{2}}\right), & r\leq r_{Coul}\\
		\frac{1}{r}, & r>r_{Coul}
	\end{cases} \label{eq:coulomb}
\end{equation}
where $ \alpha_{f} $ is the fine structure constant and $r_{Coul} = 1.47$ fm is the Coulomb radius.
%%%%%%%%%%%
		\section{Numerical Results} \label{sec:result}		
		As it has been shown in~\cite{hiyama2008prc}, the  $ _{\Xi}^{5}H $ is not bound with Nijmegen ND and ESC04 $ \Xi N$ potential, actually it was found to be a Coulomb-assisted bound state. While recently Le et al. ~\cite{le2021} reported a clearly bound state about $ 2.16 \pm 0.1 $ MeV employing a chiral $  \Xi N$ interactions. 		
		In Fig~\ref{fig:NXi_pot}, I draw (a) ESC08c  and (b) HAL QCD potentials in $^{11}S_{0},^{31}S_{0},^{13}S_{1}$
		and $^{33}S_{1}$ channels and their corresponding  spin- isospin averaged potential $ \bar{V}_{\Xi N} $.
		 Also, $ \Xi N$ phase shifts in each channels relevant to above potentials are drawn and given in Fig. 2 of Ref~\cite{Hiyama2020}. It is seen that these two potentials are different.		
 It is therefore desirable to examine the $ _{\Xi}^{5}H $ by ESC08c  and HAL QCD potentials for the purpose of seeing whether such  differences are manifest in the predictions for low energy properties.
 Since the $ \alpha $ particle is strongly bound, the mass gap between $\Xi N$ and $\Lambda \Lambda$ is somehow abolished~\cite{10.1143/ptp.117.251} that  makes this light hypernucleus to be interesting and to be studied straightforwardly. 
 
\begin{figure*}[hbt!]
			\centering
			\includegraphics[scale=1.0]{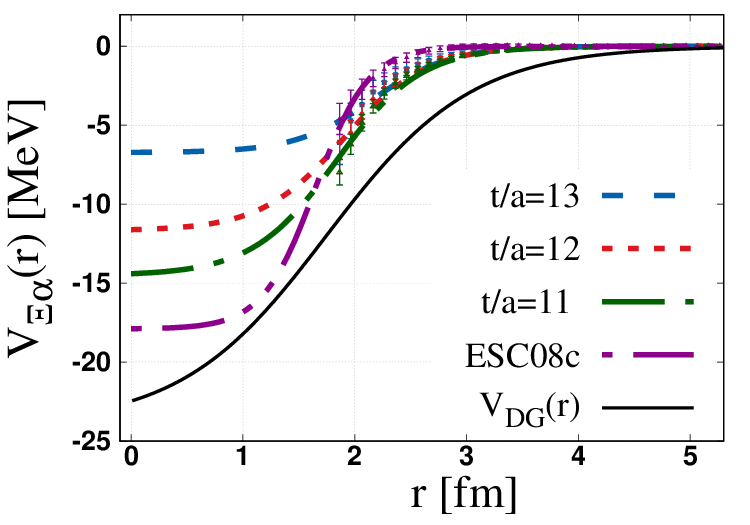}
			\caption{$V_{\alpha\Xi}\left(r\right)$, the single-folding potentials that have been obtained by Eq.~\eqref{eq:V_alfaOmega}  
				using  $\Xi N $ potential models of DG (solid black line),  ESC08c (dash-double-dotted magenta line) and HAL QCD at $ t/a=11 $ (dash-dotted green line), $ t/a=12 $ (dotted red line) and $ t/a=13 $ (dashed blue line). The corresponding $\Xi N $ potentials are depicted in Fig.~\ref{fig:NXi_pot}.				
				  \label{fig:vc-Xi-fit-DG}}
		\end{figure*}

$\alpha \Xi$ potential is obtained by solving Eq.~\eqref{eq:V_alfaOmega} and the resultant potentials are depicted in Fig.~\ref{fig:vc-Xi-fit-DG} at the imaginary-time distances $ t/a = 11, 12, 13 $ for HAL QCD potentials. Also for comparison the ESC08c and the common DG potentials are shown.
		%%%%%%%%%%%%%%%%
		%%%%%%%%%%%%%%%%	

\begin{figure*}[hbt!]
			\centering
			\includegraphics[scale=1.0]{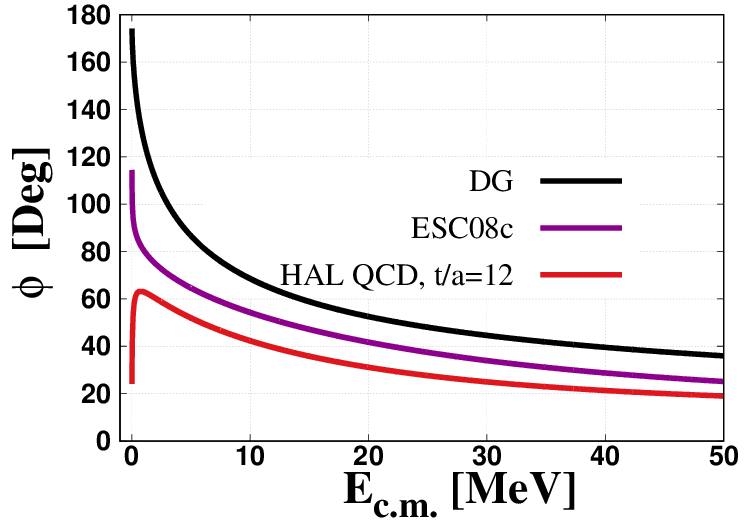}
			\caption{ $\alpha \Xi$ phase shifts using the DG~\cite{dover1983}, ESC08c and HAL QCD 
				 $ \left(t/a=12\right) $ potentials.  
				\label{fig:phase_DG_HAL}}
		\end{figure*}

		For phenomenological application and calculation of observables, such as scattering phase shifts and binding energies, I fit $V_{\alpha\Xi}\left(r\right)$ to a Wood-Saxon form using the function that given by Eq.~\eqref{eq:DG} (motivated by common DG model of potential~\cite{dover1983} ) with three parameters,  $ V_{0}, R$ and $c$,			
		\begin{equation}
			V_{\alpha \Xi}^{DG}\left(r\right)=-V_{0}\left[1+\exp\left(\frac{r-R}{c}\right)\right]^{-1} , \label{eq:DG}
		\end{equation}		
		where $ V_{0}$ is known as the depth parameter, $ R = 1.1 A^{1/3}$ with $ A $ being the mass number of the nuclear core, i.e $ \alpha $ ($ A = 4 $) and $ c $ is known as the surface diffuseness. 
		%The values of these three parameters in DG model are taken directly from Ref.~\cite{dover1983} and given in Table~\ref{tab:ERE-DO-SFP}. 
		By using the fit functions (solid lines in Fig~\ref{fig:vc-Xi-fit-DG}) as input  the Schr\"{o}dinger equation were solved in the infinite volume to extract binding energy and  scattering observables from the asymptotic behavior of the wave function.
		Fig.~\ref{fig:phase_DG_HAL} shows $\alpha \Xi$ phase shifts calculated with  DG~\cite{dover1983}, ESC08c  and HAL QCD ($ t/a=12 $) potentials for comparison. The phase shift in the case of DG potential shows an attractive interaction
		even to form a bound
		state with the binding energy about $ 2 $ MeV, while attraction in the case of ESC08c is relatively weaker, 
		and for HAL QCD potential is more weak than the previous two cases. 
				
		%%%%%%%%%%%%%%%%
		%%%%%%%%%%%%%%%%
		
		Low-energy part of $\alpha \Xi$ phase shifts in Fig.~\ref{fig:phase_DG_HAL} provides the scattering length $\left(a_{0}\right)$ and the effective range $\left( r_{0}\right)$ by employing the effective range expansion (ERE) formula up to the next-leading-order (NLO),		
		\begin{equation}
			k\cot\delta_{0}=-\frac{1}{a_{0}}+\frac{1}{2}r_{0}k^{2}+\mathcal{O}\left(k^{4}\right).\label{eq:ERE}
		\end{equation}		
			
The fit parameters,   scattering length, effective range and binding energy $B_{\alpha\Xi^{0}}$, with DG, ESC08c and HAL QCD potentials are given in Table~\ref{tab:ERE-DO-SFP}. The fit function by these parameters are plotted in Fig.~\ref{fig:vc-Xi-fit-DG} by solid lines. 
%Also, $V_{\alpha \Xi}^{DG}\left(r\right)$  are shown for comparison. 
As quoted in the caption of Table~\ref{tab:ERE-DO-SFP}, the numbers between parenthesis correspond to the calculations by using $ \Xi $  mass derived by the lattice simulations~\cite{sasaki2020}, where they are slightly bigger than experimental mass. Since by increasing the mass, contribution of repulsive
kinetic energy will decrease and finally leads to slightly deeper binding energies.
 Moreover,  binding energies,  $B_{\alpha\Xi^{-}}$ $ \left(B_{\alpha\Xi^{0}}\right) $, with (without) Coulomb interactions are given.
  A Coulomb potential due to a uniformly charged sphere of radius $ r_{Coul} $ is included, see Eq.~\eqref{eq:coulomb}.  
If a nuclear-unbound system 
 turn into bound state with support of the attractive Coulomb potential, that is called a Coulomb-assisted bound state. 
  According to behaviour of phase shifts  in Fig.~\ref{fig:phase_DG_HAL} and data in Table~\ref{tab:ERE-DO-SFP}  none of the ESC08c potential and  the HAL QCD potentials ($ t/a=11, 12 $ and $ t/a=13 $) supports a bound states for $ _{\Xi}^{5}H $.
  Nevertheless, if I switch on the Coulomb interaction, the $ _{\Xi}^{5}H $ system become  Coulomb-assisted bound state and the corresponding binding energies $ B_{\alpha\Xi^{-}} $ are given in Table~\ref{tab:ERE-DO-SFP}. Definitely as expected, their values are small which is less than $ 1 $ MeV.  Moreover, when the strong interactions are switched off no bound or resonance states are obtained. So, $ _{\Xi}^{5}H $ cannot be a Coulomb-bound (atomic) state. 
  %These results imply that $\alpha \Xi$ potential is shallow.
% atomic states which are  (almost) purely bound by the Coulomb interaction.
%Of course, Coulomb bound (Ξ− -
%atomic) states are obtained, even if the strong interactions are switched off.

Based on Hiyama et al.'s numerical results~\cite{Hiyama2020}, it is expected that the differences in magnitude of binding energies between the ESC08c and HAL QCD potentials should be almost significant, but it is not large here. 
In addition to the fact that my methods are different, there are two reasons.
One reason is that I have used the spin- and isospin averaged potentials here but they did not. Another reason could be that Hiyama et al., in order to treat the coupled-channel effects, they normalized contribution from higher channels $ \Xi N-\Lambda\Sigma-\Sigma\Sigma $ in $ ^{33}S_{1} $  into a $ \Xi N-\Xi N $ effective central potential by including a Gaussian function as,
 $ -233\exp\left(-r^{2}\right) $~\cite{Hiyama2020}. This factor dramatically increases the attractive power of the potential in  ESC08 case. I did not inserted this factor in my calculations. 

%%%%%%%%%%%%%%%%
\begin{table}[hbt!]
\caption{
	The fit parameters of $\alpha \Xi$ potential in Eq.~\eqref{eq:DG}
	and the corresponding low-energy parameter, scattering length $ a_{0} $, effective range $r_{0}$ and binding energy $B_{\alpha\Xi^{0}}$, are given for DG, ESC08c and HAL QCD potentials.
	Here the NB is an acronym for No Bound or resonance states.
	 The parameter of DG $\alpha \Xi$ potentials are taken directly from Ref.~\cite{dover1983}.
The results have been achieved by using the experimental masses of $\alpha$ and $\Xi$, $3727.38\:\textrm{MeV}/c^{2}$ and $1318.07\:\textrm{MeV}/c^{2}$ respectively. Also, the results corresponding to $  \Xi $ mass value derived by the HAL QCD Collaboration $ 1355.2 \:\textrm{MeV}/c^{2}$  are given between parenthesis. $B_{\alpha\Xi^{-}}$ $\left(B_{\alpha\Xi^{0}}\right)$ is the binding energy with (without) Coulomb potential given by Eq.~\eqref{eq:coulomb}. 
In order to get a comprehensive evaluation, the experimental ERE parameters for neutron-neutron are 
$ \left(a_{0},r_{0}\right)=\left(-18.5,2.80\right) $ fm.
\label{tab:ERE-DO-SFP}}. 	
	\begin{tabular}{cccccccc}
		\hline
		\hline 
	Model & $ V_{0} $ (MeV)& $ R $(fm)& $ c $ (fm)& $a_{0}$(fm)&$r_{0}$(fm)& $B_{\alpha\Xi^{0}}$(MeV)& $B_{\alpha\Xi^{-}}$(MeV) \\
	\hline 		
	DG        & $ 24  $& $ 1.74 $ & $ 0.65 $ & $ -4.9(-4.9)$  & $ 1.9(1.9) $   & $ 2.0(2.1) $ & $3.4(3.5)$\\
	ESC08c    & $ 18  $& $ 1.65 $ & $ 0.24 $ & $ -5.7(-5.3 )$  & $ 2.7(2.7) $   &  NB(NB)  & $0.8(0.8)$\\
	
	$ t/a=11 $& $14.5 $& $ 1.83 $ & $ 0.38 $ & $ -6.2 (-6.0)$ & $ 3.1 (3.0)$ &  NB(NB)& $0.8(0.9)$\\
%                  14.5143     2.65274     1.83519 scatt_length=-6.20854 Effective-Range=3.13146

	$ t/a=12 $& $11.7 $& $ 1.90 $ & $ 0.37 $ & $ -7.3 (-6.9)$ & $ 3.7 (3.7)$ &  NB(NB)   & $0.5(0.5)$\\
%                  11.68     2.71197     1.90043 scatt_length=-7.27804 Effective-Range=3.7328
	$ t/a=13 $& $6.7 $ & $ 2.16 $ & $ 0.35 $ & $ -6.8 (-6.6)$ & $ 5.7 (5.6)$ &  NB(NB)   & $0.2(0.2)$\\
%                  6.73362     2.88601     2.16084 catt_length=-6.81578 Effective-Range=5.74386
	\hline
	\hline 	
\end{tabular}
\end{table}
%%%%%%%%
%%%%%%%%%%%%%%%%%%%%%%%%
\section{Summary and conclusions\label{sec:Summary-and-conclusions}}
 $ _{\Xi}^{5}H $ hypernuclei is studied by several few-body technics by different $ \Xi N $ interactions and inconsistent
  results are reported. Since there are not any sufficient experimental data about $\Xi$-hypernuclei, the lattice QCD simulation is the most reliable approach. Recently, the HAL QCD Collaboration derived  $\Xi N$ interaction with quark
  masses near the physical point $ m_{\pi}\simeq146.4 $ MeV by lattice QCD simulations.
   Therefore, I examined the $ _{\Xi}^{5}H $ by using  two significantly different spin- and isospin averaged potentials, i.e   phenomenological Nijmegen ESC08c $\Xi N$ potential and first principles HAL QCD $\Xi N$ potential for the purpose of seeing whether such differences are manifest in the predictions for low energy properties.
My assumptions were as follows:  
\romannum{1}. $ \alpha $ is strongly bound and low compressible particle. 
\romannum{2}. The mass gap between $\Xi N$ and $\Lambda \Lambda$ is somehow removed and treated by coupled-channel HAL QCD method.
\romannum{3}. Coupled-channel contribution from higher channels $ \Xi N-\Lambda\Sigma-\Sigma\Sigma $ in $ ^{33}S_{1} $ is not considered.
\romannum{4}. $ _{\Xi}^{5}H $ must show an explicit core-$\Xi$ structure. 
\romannum{5}. Both $\alpha$ and $\Xi$ particles are in S-wave states.
\romannum{6}. $\Xi N$ potential is spin- and isospin averaged potential. 

  $\alpha+\Xi$ potentials were obtained by using SFP model. Then the resultant $\alpha \Xi$ potentials  were fitted by a Woods-Saxon type functions.   
The binding energies and scattering phase shift parameters were calculated by solving the
Schr\"{o}dinger equation using the fit potentials as input for DG, ESC08c and HAL QCD (at $ t/a=11,12,13 $) potentials. The numerical results showed that 
 none of the potentials support nuclear bound states for $ _{\Xi}^{5}H $. But when attractive Coulomb interaction was taken into account, the $ _{\Xi}^{5}H $ system become  Coulomb-assisted bound state around $ 0.5 $ MeV, although it was still very shallow.   
 
 Curiously enough, the results were in agreement with estimations of Hiyama et al.~\cite{hiyama2008prc} where they have found just  Coulomb-assisted bound state around  $ 0.57 $ MeV for $ k_{f}=0.9 $ (there, the binding energies was  a function of the Fermi momentum of nuclear matter, $ k_{f}$)
    and roughly consistent with the calculation by Myint and Akaishi~\cite{10.1143/ptp.117.251}. Myint and Akaishi  asserted explicitly that the large part of  $ 1.7 $ MeV binding energy is due to $\ensuremath{\Xi^{-}+\alpha}$ Coulomb interaction.

 In conclusion,  
 even though  Nijmegen ESC08c  and  HAL QCD $\Xi N$ potential have significantly dissimilar isospin  and spin components, my numerical results showed that both potential models lead to  almost qualitatively consistent results. 
 Namely, $ _{\Xi}^{5}H $ cannot be a pure atomic or nuclear bound state, instead, it could be only a Coulomb-assisted resonance state that appears about  $ 0.5 $ MeV below the threshold of $\alpha+\Xi^{-} $.
The derived $\alpha+\Xi$ potential can be used in few-body model based on $\alpha$-cluster structures of $\Xi$ hypernuclei~\cite{hiyama2018annurev,le2021}.

%\section*{Acknowledgement}
				
%%%%%%%%%%%

%\section*{References}
				
\bibliography{Refs.bib}

\end{document}